\pdfoutput=1

\documentclass[preprint, prd, tightenlines, superscriptaddress]{revtex4-1}

\usepackage{graphicx}
\usepackage{dcolumn}
\usepackage{color}
\usepackage{multirow}
\usepackage{placeins}
\usepackage{amsmath}
\usepackage{amssymb}
\usepackage{amsfonts}
\usepackage{upgreek}
\usepackage{xspace}
\usepackage{url}
\usepackage{hyperref}

\input{belle2sym.tex}

\begin{document}

\vspace*{0.5\baselineskip}
\resizebox{!}{3cm}{\includegraphics{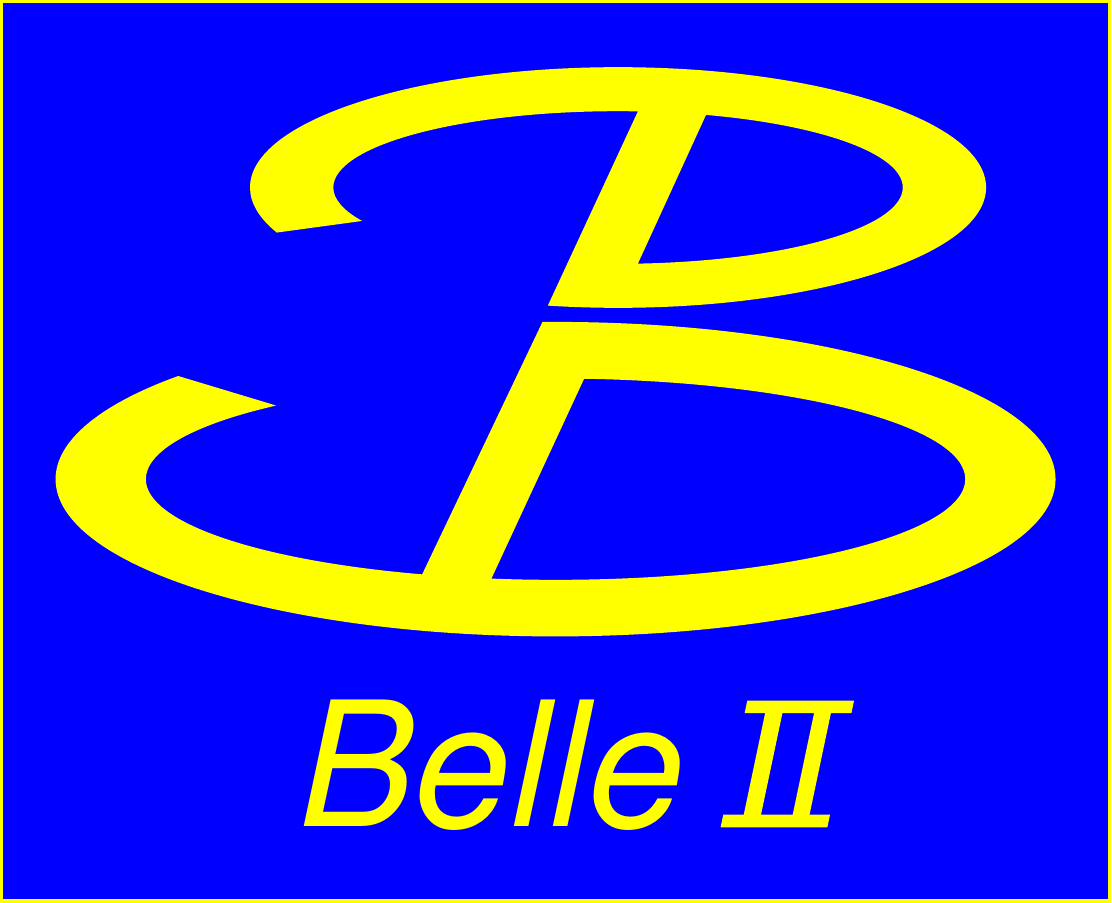}}

\vspace*{-4\baselineskip}
\begin{flushright}
BELLE2-CONF-PH-2020-006\\
\today  
\end{flushright}
\vspace*{2.5\baselineskip}

\begin{center}
{\LARGE 
Rediscovery of $B \to \phi K^{(*)}$ decays and measurement of the longitudinal
polarization fraction $f_L$ in $B \to \phi K^{*}$ decays using the Summer
2020 Belle II dataset
}\\[8pt]
\end{center}

\newcommand{\instSinica}{Academia Sinica, Taipei 11529, Taiwan}
\newcommand{\instCPPM}{Aix Marseille Universit\'{e}, CNRS/IN2P3, CPPM, 13288 Marseille, France}
\newcommand{\instBeihang}{Beihang University, Beijing 100191, China}
\newcommand{\instBUAP}{Benemerita Universidad Autonoma de Puebla, Puebla 72570, Mexico}
\newcommand{\instBNL}{Brookhaven National Laboratory, Upton, New York 11973, U.S.A.}
\newcommand{\instBINP}{Budker Institute of Nuclear Physics SB RAS, Novosibirsk 630090, Russian Federation}
\newcommand{\instCMU}{Carnegie Mellon University, Pittsburgh, Pennsylvania 15213, U.S.A.}
\newcommand{\instCinvestavIPN}{Centro de Investigacion y de Estudios Avanzados del Instituto Politecnico Nacional, Mexico City 07360, Mexico}
\newcommand{\instPrague}{Faculty of Mathematics and Physics, Charles University, 121 16 Prague, Czech Republic}
\newcommand{\instChiangMai}{Chiang Mai University, Chiang Mai 50202, Thailand}
\newcommand{\instChiba}{Chiba University, Chiba 263-8522, Japan}
\newcommand{\instChonnam}{Chonnam National University, Gwangju 61186, South Korea}
\newcommand{\instConacyt}{Consejo Nacional de Ciencia y Tecnolog\'{\i}a, Mexico City 03940, Mexico}
\newcommand{\instDESY}{Deutsches Elektronen--Synchrotron, 22607 Hamburg, Germany}
\newcommand{\instDuke}{Duke University, Durham, North Carolina 27708, U.S.A.}
\newcommand{\instITAR}{Institute of Theoretical and Applied Research (ITAR), Duy Tan University, Hanoi 100000, Vietnam}
\newcommand{\instENEA}{ENEA Casaccia, I-00123 Roma, Italy}
\newcommand{\instEri}{Earthquake Research Institute, University of Tokyo, Tokyo 113-0032, Japan}
\newcommand{\instJuelich}{Forschungszentrum J\"{u}lich, 52425 J\"{u}lich, Germany}
\newcommand{\instFuJen}{Department of Physics, Fu Jen Catholic University, Taipei 24205, Taiwan}
\newcommand{\instFudan}{Key Laboratory of Nuclear Physics and Ion-beam Application (MOE) and Institute of Modern Physics, Fudan University, Shanghai 200443, China}
\newcommand{\instGoettingen}{II. Physikalisches Institut, Georg-August-Universit\"{a}t G\"{o}ttingen, 37073 G\"{o}ttingen, Germany}
\newcommand{\instGifu}{Gifu University, Gifu 501-1193, Japan}
\newcommand{\instSOKENDAI}{The Graduate University for Advanced Studies (SOKENDAI), Hayama 240-0193, Japan}
\newcommand{\instGyeongsang}{Gyeongsang National University, Jinju 52828, South Korea}
\newcommand{\instHanyang}{Department of Physics and Institute of Natural Sciences, Hanyang University, Seoul 04763, South Korea}
\newcommand{\instKEK}{High Energy Accelerator Research Organization (KEK), Tsukuba 305-0801, Japan}
\newcommand{\instJPARC}{J-PARC Branch, KEK Theory Center, High Energy Accelerator Research Organization (KEK), Tsukuba 305-0801, Japan}
\newcommand{\instHSE}{Higher School of Economics (HSE), Moscow 101000, Russian Federation}
\newcommand{\instIISER}{Indian Institute of Science Education and Research Mohali, SAS Nagar, 140306, India}
\newcommand{\instIITBhubaneswar}{Indian Institute of Technology Bhubaneswar, Satya Nagar 751007, India}
\newcommand{\instIITGuwahati}{Indian Institute of Technology Guwahati, Assam 781039, India}
\newcommand{\instIITHyderabad}{Indian Institute of Technology Hyderabad, Telangana 502285, India}
\newcommand{\instIITMadras}{Indian Institute of Technology Madras, Chennai 600036, India}
\newcommand{\instIndiana}{Indiana University, Bloomington, Indiana 47408, U.S.A.}
\newcommand{\instIHEPRussia}{Institute for High Energy Physics, Protvino 142281, Russian Federation}
\newcommand{\instHEPHYVienna}{Institute of High Energy Physics, Vienna 1050, Austria}
\newcommand{\instIHEPChina}{Institute of High Energy Physics, Chinese Academy of Sciences, Beijing 100049, China}
\newcommand{\instChennai}{Institute of Mathematical Sciences, Chennai 600113, India}
\newcommand{\instIPP}{Institute of Particle Physics (Canada), Victoria, British Columbia V8W 2Y2, Canada}
\newcommand{\instIOP}{Institute of Physics, Vietnam Academy of Science and Technology (VAST), Hanoi, Vietnam}
\newcommand{\instIFIC}{Instituto de Fisica Corpuscular, Paterna 46980, Spain}
\newcommand{\instFrascati}{INFN Laboratori Nazionali di Frascati, I-00044 Frascati, Italy}
\newcommand{\instNapoliINFN}{INFN Sezione di Napoli, I-80126 Napoli, Italy}
\newcommand{\instPadovaINFN}{INFN Sezione di Padova, I-35131 Padova, Italy}
\newcommand{\instPerugiaINFN}{INFN Sezione di Perugia, I-06123 Perugia, Italy}
\newcommand{\instPisaINFN}{INFN Sezione di Pisa, I-56127 Pisa, Italy}
\newcommand{\instRomaINFN}{INFN Sezione di Roma, I-00185 Roma, Italy}
\newcommand{\instRomaTreINFN}{INFN Sezione di Roma Tre, I-00146 Roma, Italy}
\newcommand{\instTorinoINFN}{INFN Sezione di Torino, I-10125 Torino, Italy}
\newcommand{\instTriesteINFN}{INFN Sezione di Trieste, I-34127 Trieste, Italy}
\newcommand{\instJAEA}{Advanced Science Research Center, Japan Atomic Energy Agency, Naka 319-1195, Japan}
\newcommand{\instMainz}{Johannes Gutenberg-Universit\"{a}t Mainz, Institut f\"{u}r Kernphysik, D-55099 Mainz, Germany}
\newcommand{\instGiessen}{Justus-Liebig-Universit\"{a}t Gie\ss{}en, 35392 Gie\ss{}en, Germany}
\newcommand{\instKarlsruhe}{Institut f\"{u}r Experimentelle Teilchenphysik, Karlsruher Institut f\"{u}r Technologie, 76131 Karlsruhe, Germany}
\newcommand{\instKennesaw}{Kennesaw State University, Kennesaw, Georgia 30144, U.S.A.}
\newcommand{\instKitasato}{Kitasato University, Sagamihara 252-0373, Japan}
\newcommand{\instKISTI}{Korea Institute of Science and Technology Information, Daejeon 34141, South Korea}
\newcommand{\instKorea}{Korea University, Seoul 02841, South Korea}
\newcommand{\instKSU}{Kyoto Sangyo University, Kyoto 603-8555, Japan}
\newcommand{\instKyotoU}{Kyoto University, Kyoto 606-8501, Japan}
\newcommand{\instKyungpook}{Kyungpook National University, Daegu 41566, South Korea}
\newcommand{\instLPI}{P.N. Lebedev Physical Institute of the Russian Academy of Sciences, Moscow 119991, Russian Federation}
\newcommand{\instLNNU}{Liaoning Normal University, Dalian 116029, China}
\newcommand{\instLMU}{Ludwig Maximilians University, 80539 Munich, Germany}
\newcommand{\instLuther}{Luther College, Decorah, Iowa 52101, U.S.A.}
\newcommand{\instMNITJaipur}{Malaviya National Institute of Technology Jaipur, Jaipur 302017, India}
\newcommand{\instMPP}{Max-Planck-Institut f\"{u}r Physik, 80805 M\"{u}nchen, Germany}
\newcommand{\instMPGHLL}{Semiconductor Laboratory of the Max Planck Society, 81739 M\"{u}nchen, Germany}
\newcommand{\instMcGill}{McGill University, Montr\'{e}al, Qu\'{e}bec, H3A 2T8, Canada}
\newcommand{\instMETU}{Middle East Technical University, 06531 Ankara, Turkey}
\newcommand{\instMEPhI}{Moscow Physical Engineering Institute, Moscow 115409, Russian Federation}
\newcommand{\instNagoya}{Graduate School of Science, Nagoya University, Nagoya 464-8602, Japan}
\newcommand{\instNagoyaKMI}{Kobayashi-Maskawa Institute, Nagoya University, Nagoya 464-8602, Japan}
\newcommand{\instNagoyaIAR}{Institute for Advanced Research, Nagoya University, Nagoya 464-8602, Japan}
\newcommand{\instNaraWu}{Nara Women's University, Nara 630-8506, Japan}
\newcommand{\instUNAM}{National Autonomous University of Mexico, Mexico City, Mexico}
\newcommand{\instNTUTaiwan}{Department of Physics, National Taiwan University, Taipei 10617, Taiwan}
\newcommand{\instNUUTaiwan}{National United University, Miao Li 36003, Taiwan}
\newcommand{\instKrakow}{H. Niewodniczanski Institute of Nuclear Physics, Krakow 31-342, Poland}
\newcommand{\instNiigata}{Niigata University, Niigata 950-2181, Japan}
\newcommand{\instNSU}{Novosibirsk State University, Novosibirsk 630090, Russian Federation}
\newcommand{\instOkinawa}{Okinawa Institute of Science and Technology, Okinawa 904-0495, Japan}
\newcommand{\instOsakaCity}{Osaka City University, Osaka 558-8585, Japan}
\newcommand{\instRCNP}{Research Center for Nuclear Physics, Osaka University, Osaka 567-0047, Japan}
\newcommand{\instPNNL}{Pacific Northwest National Laboratory, Richland, Washington 99352, U.S.A.}
\newcommand{\instPanjab}{Panjab University, Chandigarh 160014, India}
\newcommand{\instPeking}{Peking University, Beijing 100871, China}
\newcommand{\instPanjabPAU}{Punjab Agricultural University, Ludhiana 141004, India}
\newcommand{\instRIKENMSL}{Meson Science Laboratory, Cluster for Pioneering Research, RIKEN, Saitama 351-0198, Japan}
\newcommand{\instRIKEN}{Theoretical Research Division, Nishina Center, RIKEN, Saitama 351-0198, Japan}
\newcommand{\instXavier}{St. Francis Xavier University, Antigonish, Nova Scotia, B2G 2W5, Canada}
\newcommand{\instSeoul}{Seoul National University, Seoul 08826, South Korea}
\newcommand{\instShandong}{Shandong University, Jinan 250100, China}
\newcommand{\instSPU}{Showa Pharmaceutical University, Tokyo 194-8543, Japan}
\newcommand{\instSoochow}{Soochow University, Suzhou 215006, China}
\newcommand{\instSoongsil}{Soongsil University, Seoul 06978, South Korea}
\newcommand{\instLjubljanaJSI}{J. Stefan Institute, 1000 Ljubljana, Slovenia}
\newcommand{\instKyiv}{Taras Shevchenko National Univ. of Kiev, Kiev, Ukraine}
\newcommand{\instTata}{Tata Institute of Fundamental Research, Mumbai 400005, India}
\newcommand{\instTUM}{Department of Physics, Technische Universit\"{a}t M\"{u}nchen, 85748 Garching, Germany}
\newcommand{\instECUTUM}{Excellence Cluster Universe, Technische Universit\"{a}t M\"{u}nchen, 85748 Garching, Germany}
\newcommand{\instTelAviv}{Tel Aviv University, School of Physics and Astronomy, Tel Aviv, 69978, Israel}
\newcommand{\instToho}{Toho University, Funabashi 274-8510, Japan}
\newcommand{\instTohoku}{Department of Physics, Tohoku University, Sendai 980-8578, Japan}
\newcommand{\instTitech}{Tokyo Institute of Technology, Tokyo 152-8550, Japan}
\newcommand{\instTokyoMetropolitan}{Tokyo Metropolitan University, Tokyo 192-0397, Japan}
\newcommand{\instUAS}{Universidad Autonoma de Sinaloa, Sinaloa 80000, Mexico}
\newcommand{\instNapoliUNIV}{Dipartimento di Scienze Fisiche, Universit\`{a} di Napoli Federico II, I-80126 Napoli, Italy}
\newcommand{\instNapoliUNIVA}{Dipartimento di Agraria, Universit\`{a} di Napoli Federico II, I-80055 Portici (NA), Italy}
\newcommand{\instPadovaUNIV}{Dipartimento di Fisica e Astronomia, Universit\`{a} di Padova, I-35131 Padova, Italy}
\newcommand{\instPerugiaUNIV}{Dipartimento di Fisica, Universit\`{a} di Perugia, I-06123 Perugia, Italy}
\newcommand{\instPisaUNIV}{Dipartimento di Fisica, Universit\`{a} di Pisa, I-56127 Pisa, Italy}
\newcommand{\instRomaUNIV}{Universit\`{a} di Roma ``La Sapienza,'' I-00185 Roma, Italy}
\newcommand{\instRomaTreUNIV}{Dipartimento di Matematica e Fisica, Universit\`{a} di Roma Tre, I-00146 Roma, Italy}
\newcommand{\instTorinoUNIV}{Dipartimento di Fisica, Universit\`{a} di Torino, I-10125 Torino, Italy}
\newcommand{\instTriesteUNIV}{Dipartimento di Fisica, Universit\`{a} di Trieste, I-34127 Trieste, Italy}
\newcommand{\instMontreal}{Universit\'{e} de Montr\'{e}al, Physique des Particules, Montr\'{e}al, Qu\'{e}bec, H3C 3J7, Canada}
\newcommand{\instIJCLab}{Universit\'{e} Paris-Saclay, CNRS/IN2P3, IJCLab, 91405 Orsay, France}
\newcommand{\instIPHC}{Universit\'{e} de Strasbourg, CNRS, IPHC, UMR 7178, 67037 Strasbourg, France}
\newcommand{\instAdelaide}{Department of Physics, University of Adelaide, Adelaide, South Australia 5005, Australia}
\newcommand{\instBonn}{University of Bonn, 53115 Bonn, Germany}
\newcommand{\instUBC}{University of British Columbia, Vancouver, British Columbia, V6T 1Z1, Canada}
\newcommand{\instCincinnati}{University of Cincinnati, Cincinnati, Ohio 45221, U.S.A.}
\newcommand{\instFlorida}{University of Florida, Gainesville, Florida 32611, U.S.A.}
\newcommand{\instHamburg}{University of Hamburg, 20148 Hamburg, Germany}
\newcommand{\instHawaii}{University of Hawaii, Honolulu, Hawaii 96822, U.S.A.}
\newcommand{\instHeidelberg}{University of Heidelberg, 68131 Mannheim, Germany}
\newcommand{\instLjubljanaUniLJ}{Faculty of Mathematics and Physics, University of Ljubljana, 1000 Ljubljana, Slovenia}
\newcommand{\instLouisville}{University of Louisville, Louisville, Kentucky 40292, U.S.A.}
\newcommand{\instMalaya}{National Centre for Particle Physics, University Malaya, 50603 Kuala Lumpur, Malaysia}
\newcommand{\instLjubljanaUM}{University of Maribor, 2000 Maribor, Slovenia}
\newcommand{\instMelbourne}{School of Physics, University of Melbourne, Victoria 3010, Australia}
\newcommand{\instMississippi}{University of Mississippi, University, Mississippi 38677, U.S.A.}
\newcommand{\instUOM}{University of Miyazaki, Miyazaki 889-2192, Japan}
\newcommand{\instNovaGorica}{University of Nova Gorica, 5000 Nova Gorica, Slovenia}
\newcommand{\instPittsburgh}{University of Pittsburgh, Pittsburgh, Pennsylvania 15260, U.S.A.}
\newcommand{\instUSTC}{University of Science and Technology of China, Hefei 230026, China}
\newcommand{\instSAlabama}{University of South Alabama, Mobile, Alabama 36688, U.S.A.}
\newcommand{\instSCarolina}{University of South Carolina, Columbia, South Carolina 29208, U.S.A.}
\newcommand{\instSydney}{School of Physics, University of Sydney, New South Wales 2006, Australia}
\newcommand{\instTabuk}{Department of Physics, Faculty of Science, University of Tabuk, Tabuk 71451, Saudi Arabia}
\newcommand{\instUTokyo}{Department of Physics, University of Tokyo, Tokyo 113-0033, Japan}
\newcommand{\instIPMU}{Kavli Institute for the Physics and Mathematics of the Universe (WPI), University of Tokyo, Kashiwa 277-8583, Japan}
\newcommand{\instVictoria}{University of Victoria, Victoria, British Columbia, V8W 3P6, Canada}
\newcommand{\instVPI}{Virginia Polytechnic Institute and State University, Blacksburg, Virginia 24061, U.S.A.}
\newcommand{\instWayneState}{Wayne State University, Detroit, Michigan 48202, U.S.A.}
\newcommand{\instYamagata}{Yamagata University, Yamagata 990-8560, Japan}
\newcommand{\instYerevan}{Alikhanyan National Science Laboratory, Yerevan 0036, Armenia}
\newcommand{\instYonsei}{Yonsei University, Seoul 03722, South Korea}
\affiliation{\instCPPM}
\affiliation{\instBeihang}
\affiliation{\instBNL}
\affiliation{\instBINP}
\affiliation{\instCMU}
\affiliation{\instCinvestavIPN}
\affiliation{\instPrague}
\affiliation{\instChiangMai}
\affiliation{\instChiba}
\affiliation{\instChonnam}
\affiliation{\instConacyt}
\affiliation{\instDESY}
\affiliation{\instDuke}
\affiliation{\instITAR}
\affiliation{\instEri}
\affiliation{\instJuelich}
\affiliation{\instFuJen}
\affiliation{\instFudan}
\affiliation{\instGoettingen}
\affiliation{\instGifu}
\affiliation{\instSOKENDAI}
\affiliation{\instGyeongsang}
\affiliation{\instHanyang}
\affiliation{\instKEK}
\affiliation{\instJPARC}
\affiliation{\instHSE}
\affiliation{\instIISER}
\affiliation{\instIITBhubaneswar}
\affiliation{\instIITGuwahati}
\affiliation{\instIITHyderabad}
\affiliation{\instIITMadras}
\affiliation{\instIndiana}
\affiliation{\instIHEPRussia}
\affiliation{\instHEPHYVienna}
\affiliation{\instIHEPChina}
\affiliation{\instIPP}
\affiliation{\instIOP}
\affiliation{\instIFIC}
\affiliation{\instFrascati}
\affiliation{\instNapoliINFN}
\affiliation{\instPadovaINFN}
\affiliation{\instPerugiaINFN}
\affiliation{\instPisaINFN}
\affiliation{\instRomaINFN}
\affiliation{\instRomaTreINFN}
\affiliation{\instTorinoINFN}
\affiliation{\instTriesteINFN}
\affiliation{\instJAEA}
\affiliation{\instMainz}
\affiliation{\instGiessen}
\affiliation{\instKarlsruhe}
\affiliation{\instKitasato}
\affiliation{\instKISTI}
\affiliation{\instKorea}
\affiliation{\instKSU}
\affiliation{\instKyungpook}
\affiliation{\instLPI}
\affiliation{\instLNNU}
\affiliation{\instLMU}
\affiliation{\instLuther}
\affiliation{\instMNITJaipur}
\affiliation{\instMPP}
\affiliation{\instMPGHLL}
\affiliation{\instMcGill}
\affiliation{\instMEPhI}
\affiliation{\instNagoya}
\affiliation{\instNagoyaKMI}
\affiliation{\instNagoyaIAR}
\affiliation{\instNaraWu}
\affiliation{\instNTUTaiwan}
\affiliation{\instNUUTaiwan}
\affiliation{\instKrakow}
\affiliation{\instNiigata}
\affiliation{\instNSU}
\affiliation{\instOkinawa}
\affiliation{\instOsakaCity}
\affiliation{\instRCNP}
\affiliation{\instPNNL}
\affiliation{\instPanjab}
\affiliation{\instPeking}
\affiliation{\instPanjabPAU}
\affiliation{\instRIKENMSL}
\affiliation{\instSeoul}
\affiliation{\instSPU}
\affiliation{\instSoochow}
\affiliation{\instSoongsil}
\affiliation{\instLjubljanaJSI}
\affiliation{\instKyiv}
\affiliation{\instTata}
\affiliation{\instTUM}
\affiliation{\instTelAviv}
\affiliation{\instToho}
\affiliation{\instTohoku}
\affiliation{\instTitech}
\affiliation{\instTokyoMetropolitan}
\affiliation{\instUAS}
\affiliation{\instNapoliUNIV}
\affiliation{\instPadovaUNIV}
\affiliation{\instPerugiaUNIV}
\affiliation{\instPisaUNIV}
\affiliation{\instRomaUNIV}
\affiliation{\instRomaTreUNIV}
\affiliation{\instTorinoUNIV}
\affiliation{\instTriesteUNIV}
\affiliation{\instMontreal}
\affiliation{\instIJCLab}
\affiliation{\instIPHC}
\affiliation{\instAdelaide}
\affiliation{\instBonn}
\affiliation{\instUBC}
\affiliation{\instCincinnati}
\affiliation{\instFlorida}
\affiliation{\instHawaii}
\affiliation{\instHeidelberg}
\affiliation{\instLjubljanaUniLJ}
\affiliation{\instLouisville}
\affiliation{\instMalaya}
\affiliation{\instLjubljanaUM}
\affiliation{\instMelbourne}
\affiliation{\instMississippi}
\affiliation{\instUOM}
\affiliation{\instPittsburgh}
\affiliation{\instUSTC}
\affiliation{\instSAlabama}
\affiliation{\instSCarolina}
\affiliation{\instSydney}
\affiliation{\instUTokyo}
\affiliation{\instIPMU}
\affiliation{\instVictoria}
\affiliation{\instVPI}
\affiliation{\instWayneState}
\affiliation{\instYamagata}
\affiliation{\instYerevan}
\affiliation{\instYonsei}
  \author{F.~Abudin{\'e}n}\affiliation{\instTriesteINFN} 
  \author{I.~Adachi}\affiliation{\instKEK}\affiliation{\instSOKENDAI} 
  \author{R.~Adak}\affiliation{\instFudan} 
  \author{K.~Adamczyk}\affiliation{\instKrakow} 
  \author{P.~Ahlburg}\affiliation{\instBonn} 
  \author{J.~K.~Ahn}\affiliation{\instKorea} 
  \author{H.~Aihara}\affiliation{\instUTokyo} 
  \author{N.~Akopov}\affiliation{\instYerevan} 
  \author{A.~Aloisio}\affiliation{\instNapoliUNIV}\affiliation{\instNapoliINFN} 
  \author{F.~Ameli}\affiliation{\instRomaINFN} 
% ...
\collaboration{Belle II Collaboration}

\vspace*{5mm}

\begin{abstract}

We utilize a sample of 34.6\invfb, collected by the Belle II experiment at the SuperKEKB
asymmetric energy \epem\ collider, to search for the $\Bp \to \phi \Kp$, $\Bp \to \phi \Kstarp$,
$\Bz \to \phi \KS$, and $\Bz \to \phi \Kstarz$ decays. Charmless hadronic \B\ decays
represent an important part of the Belle II physics program, and are an ideal
benchmark to test the detector capabilities in terms of tracking efficiency,
charged particle identification, vertexing, and advanced analysis techniques.
Each channel is observed with a significance that exceeds 5 standard deviations,
and we obtain measurements of their branching ratios that are in good agreement with
the world averages.
For the $B \to \phi \Kstar$ modes, we also perform a measurement of the
longitudinal polarization fraction $f_L$.

\end{abstract}

\maketitle

\section{Introduction}
\label{sec:introduction}

The $\Bz \to \phi \Kz$ channel is one of the most interesting among the charmless hadronic $B$ decays,
as it proceeds dominantly through penguin amplitudes, and is theoretically well understood~\cite{Kou:2018nap}.
The time dependent $CP$ violation analysis of this mode may reveal effects of physics beyond
the standard model, in case some significant deviation (from the tree dominated $\Bz \to J/\psi \Kz$)
is observed.

The size of the dataset collected so far by the Belle II experiment does not yet allow for such an
analysis, so as a preparatory work we focus on the rediscovery of this and of the isospin related
$\Bp \to \phi \Kp$ mode. The relevance of this work consists in the fact that these decays have
branching fractions below $10^{-5}$ and suffer from relatively high combinatorial backgrounds, mostly
arising from random combination of particles in \emph{continuum} $\epem \to q\overline{q}$
($q = u, d, s, c$) events. The rediscovery of these modes thus constitutes an important benchmark
for assessing the performance of the detector in terms of tracking efficiency, charged particle
identification, vertexing, reconstruction of intermediate resonances, and advanced analysis
techniques. The suppression of the continuum background relies on multivariate binary discriminators
and the extraction of the signal yield is performed through a multidimensional extended maximum
likelihood fit.

The inclusion of the vector-vector $B \to \phi \Kstar$ channels, which have similar branching
fractions, extends the scope of the analysis and allows for a significant measurement of the
longitudinal polarization fraction $f_L$. In the early 2000's, this quantity attracted significant
interest (the so-called \emph{polarization puzzle}) as many $B$ decays to pairs of vector mesons that
proceed predominantly through penguin amplitudes have been observed to have sizable transverse polarization,
contrary to na\"{i}ve predictions based on helicity arguments, which predict $f_L \sim 1$ (see e.g.
the section \emph{Polarization in B decays} in \cite{Tanabashi:2018oca}). The general
consensus nowadays is that the polarization puzzle can be explained satisfactorily without invoking effects
of physics beyond the standard model (for example by postulating large contributions from penguin annihilation
\cite{Kagan:2004uw} or electroweak penguin \cite{Beneke:2005we} diagrams). Still, the measurement of the
polarization is very interesting for us as it is very sensitive to effects produced by the non-uniform
detector acceptance; demonstrating the capability of producing an accurate measurement is another important
milestone for the experiment.

\section{The Belle II detector and dataset}
\label{sec:dataset}

The Belle II detector is described in detail in Ref.~\cite{Abe:2010gxa}. The innermost sub-detector is the
vertex detector (VXD), devoted to tracking and vertexing, which is comprised of two layers of silicon pixel
sensors surrounded by four layers of silicon strips. The main tracking device is a small-cell, helium ethane
based, central drift chamber (CDC), which precisely measures the momenta of charged particles and their
specific energy loss due to ionization ($dE/dx$).
Additional particle identification (PID) is provided by two Cherenkov detectors: the Time Of Propagation (TOP)
counter in the barrel region, and the Aerogel Ring Imaging CHerenkov (ARICH), which covers the forward
endcap region. Photon detection and electron identification are provided by a CsI(Tl) electromagnetic
calorimeter (ECL). All these subdetectors operate in a 1.5T  magnetic field generated by a superconducting
solenoid. The axis of the solenoid defines the $z$ axis of the laboratory reference frame, and its positive
direction coincides approximately with the momentum of the electron beam.
The iron return yoke, instrumented with scintillator strips and resistive-plate chambers,
constitutes the KLM, the sub-detector devoted to the identification of muons and $K_L$ mesons.

Operations with the complete Belle II detector at the SuperKEKB asymmetric energy \epem\ collider~\cite{Akai:2018mbz}
began in March 2019. We utilize the data collected until the first half of May 2020 at the center-of-mass (CM) energy
corresponding to the mass of the \FourS\ resonance. The sample has an integrated luminosity of 34.6\invfb,
which corresponds to 19.7 million \BpBm\ and 18.7 million \BzBzb\ pairs.
We derived the above numbers by taking the $\epem \to \FourS$ cross-section ($1.110 \pm 0.008$) nb~\cite{Bevan:2014iga},
assuming that the \FourS\ decays exclusively to \BB\ pairs, and $f_{00} = 0.487 \pm 0.010 \pm 0.08$ \cite{Aubert:2005bq},
where $f_{00}$ is the branching fraction of $\FourS \to \BzBzb$.

\section{Event selection and analysis variables}
\label{sec:selection}

We search for the final states $\Bp \to \phi \Kp$, $\Bp \to \phi \Kstarp$, $\Bz \to \phi \KS$,
and $\Bz \to \phi \Kstarz$, with $\phi \to \Kp\Km$, $\Kstarz \to \Kp\pim$, $\Kstarp \to \KS\pip$, 
and $\KS \to \pip\pim$. Unless otherwise stated, charge conjugation is always implied.

Signal candidates are selected by applying the following criteria. Charged tracks expected to
originate from the interaction point (thus excluding the daughters of \KS\ candidates) are required
to have their point of closest approach within 2 cm (0.5 cm) of the measured \epem\ interaction
point along the $z$ axis (in the transverse plane). Charged kaon candidates are selected by applying a
cut on a likelihood based binary $K-\pi$ discriminator, which combines PID information from all
the subdetectors that can provide useful information. For the bachelor kaon in $\Bp \to \phi \Kp$
and for the kaon from the $\Kstarz \to \Kp\pim$, we apply a loose requirement (whose typical
efficiency is $> 90\%$ in the relevant momentum and polar angle ranges), whereas for the $\phi$
candidate reconstruction, we require that at least one of the daughter kaons satisfies a tighter
(efficiency $> 80\%$) selection.

The invariant masses of the intermediate resonances must satisfy: $1.00 < m(\phi) < 1.05$ \gevcc,
$0.485 < m(\KS) < 0.510$ \gevcc, and $0.74 < m(K\pi) < 0.94$ \gevcc\ (the latter requirement being
valid for both \Kstarp\ and \Kstarz\ candidates).

\begin{figure}[htbp]
  \begin{center}
    \includegraphics[width=12cm]{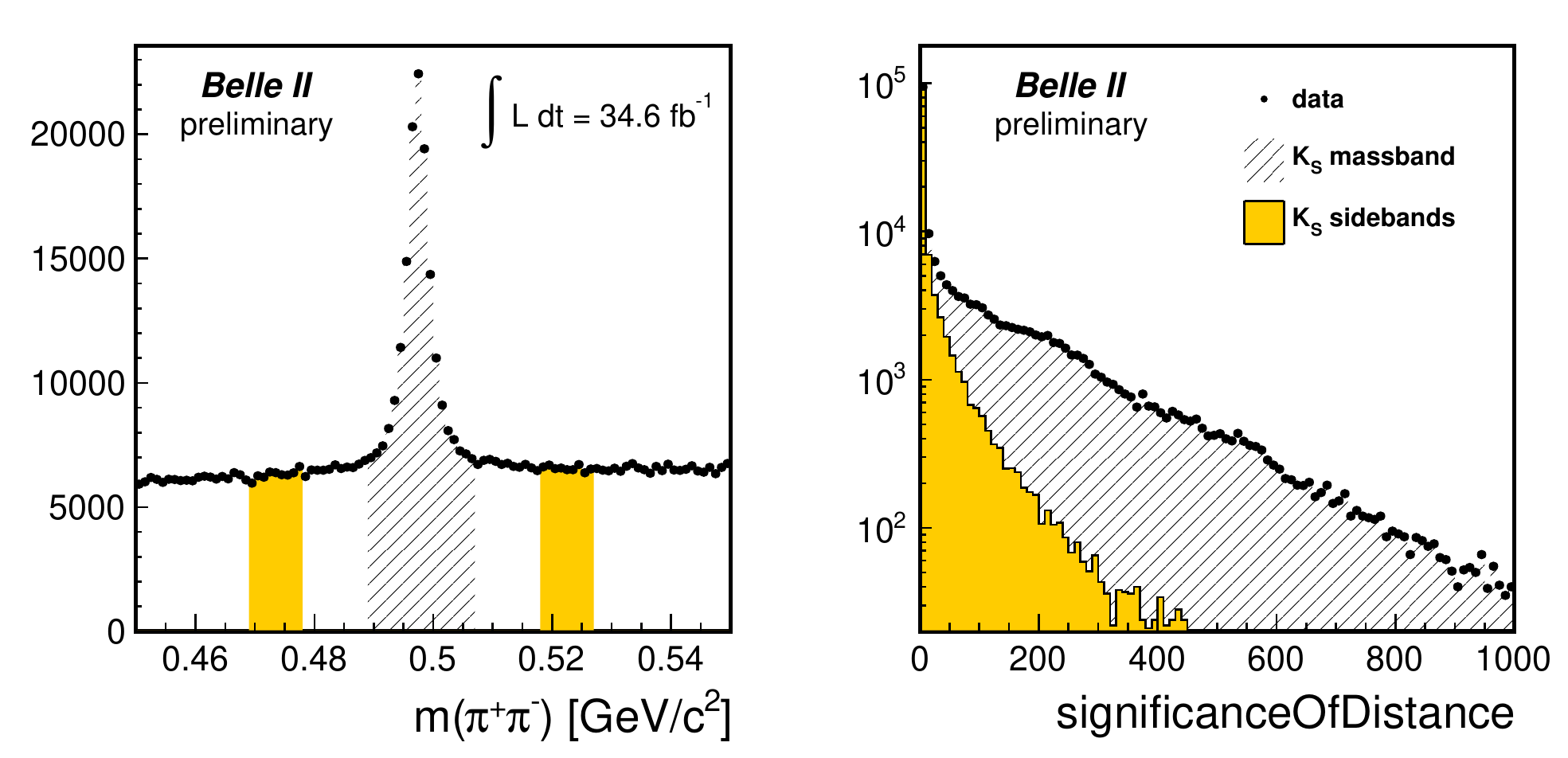} 
    \caption{Left figure: invariant mass of the \KS\ candidates, and definition of the
      massband (hatched region) and sidebands (colored region). Right: distribution
      of the significance of distance for the \KS\ massband (data points and hatched histogram)
      overlaid with the distribution taken from the sidebands (colored histogram).
      \label{fig:Ks_mass_flLenSign}}
  \end{center}
\end{figure}

To greatly enhance the purity of the \KS\ sample, we compute the \emph{significance of distance} of
each candidate, by taking the ratio between the length of the segment that connects the interaction
point with the reconstructed \KS\ decay vertex and the uncertainty in the determination of the
decay vertex. We retain candidates in which the significance of distance is greater than 50
(this requirement having an efficiency $> 80\%$). Figure~\ref{fig:Ks_mass_flLenSign} shows the
distributions of the invariant mass of the \KS\ candidates, and that of the significance of distance,
separately for genuine \KS\ candidates and random pion pair combinations.

To reduce the dominant backgrounds arising from random combinations of particles in continuum events,
we consider the ratio of the second to zeroth Fox-Wolfram moments ($R_2$) \cite{Fox:1978vu} and the
cosine of the angle between the thrust axis of the signal $B$ candidate and that of the
rest of the event (cosTBTO). We require $R_2 < 0.5$ and cosTBTO $< 0.95$. These cuts are quite loose,
to keep the signal reconstruction efficiency as high as reasonably possible. We then
combine 30 variables sensitive to the event shape and train a multivariate BDT discriminator
to distinguish between signal events (which are typically \emph{spherical}) and continuum
events (more \emph{jet-like}). The discriminator is optimized for each individual final state,
and it is utilized as one of the input variables in the final maximum likelihood fit.

As in most analyses in which the signal $B$ candidate is fully reconstructed, we use the standard
beam-constrained mass $M_{\rm bc}$ and the difference between the reconstructed and expected
$B$ candidate energies $\Delta E$:
\begin{eqnarray}
M_{\rm bc} & = & \sqrt{E^{*2}_{\rm beam} c^4 - p^{*2}_B c^2} \; , \\
\Delta E & = & E^*_B - E^*_{\rm beam} \; , 
\end{eqnarray}
where $(E^*_B, p^*_B)$ are the measured energy and momentum of the candidate $B$, and $E^*_{\rm beam} = \sqrt{s}/2$.
All quantities are calculated in the CM system. For the final fit, we require that candidates
satisfy $M_{\rm bc} > 5.25$ \gevcc\ and $|\Delta E| < 0.2$ \gev.

For $B \to V_1 V_2$ decays, where $V_i$ is a vector meson decaying to two pseudoscalar mesons, the
angular distribution, after integrating over the angle between the decay planes of $V_1$ and $V_2$,
is described by:
\begin{equation}\label{eq:helicities}
  \frac{1}{\Gamma}\frac{d^2 \Gamma}{d \cos \theta_1 d\cos \theta_2} = \frac{9}{4} \left[
    \frac{1}{4} (1 - f_L) \sin^2 \theta_1 \sin^2 \theta_2 + f_L \cos^2 \theta_1 \cos^2 \theta_2
    \right] \; ,
\end{equation}
where the subscript $L$ refers to the longitudinally polarized component,
and $f_L$ is the fraction of the longitudinally polarized component.

For the $\phi$ and $\Kstar$ resonances, the helicity angles $\theta_{H, \phi}$ and $\theta_{H,\Kstar}$
are defined as the angle between the momentum of the daughter kaon (the negatively charged in the
case of the $\phi$, the only kaon in the case of the \Kstar) and the flight direction of the
$\phi$/$\Kstar$, as measured in the $\phi$/$\Kstar$ rest frame.

The helicity angles $\theta_{H, \phi}$ and $\theta_{H,\Kstar}$ are the key variables for the measurement
of the longitudinal polarization fraction $f_L$. In the case of $B \to \phi K$, where the
spin of the $\phi$ is forced by angular momentum conservation to be perpendicular to the $\phi$
momentum, the variable $\theta_{H, \phi}$ provides additional discrimination against the continuum background
and the nonresonant $B \to \Kp \Km K$ component; for both backgrounds, the $\cos \theta_{H, \phi}$ distribution
is expected to be roughly flat.

\begin{figure}[htbp]
  \begin{center}
    \includegraphics[width=12cm]{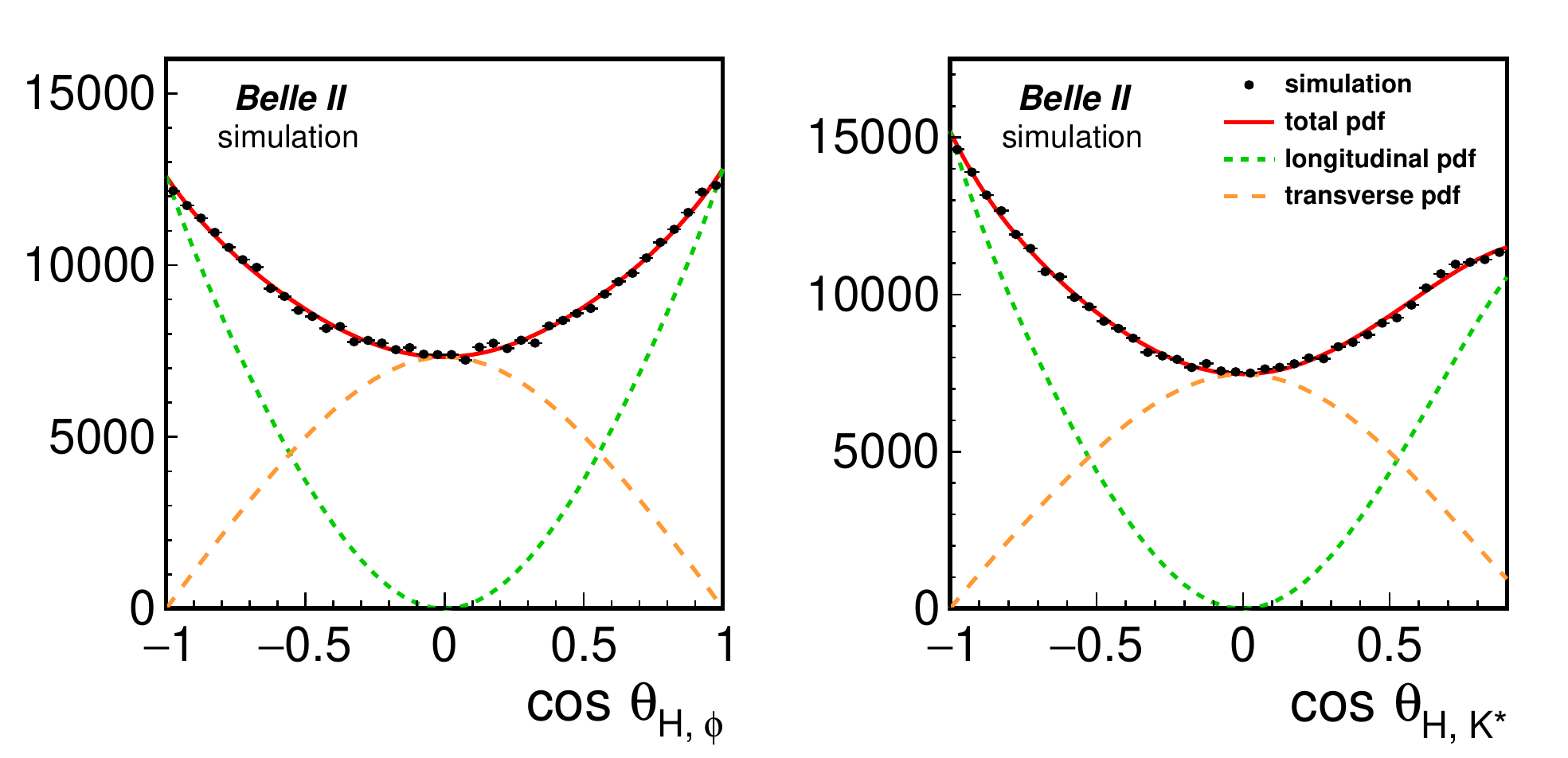} 
    \caption{Distributions of $\cos \theta_{H, \phi}$ (left) and $\cos \theta_{H,\Kstar}$ (right)
      taken from the simulation, generated with the $f_L = 0.5$ hypothesis and no background.
      The green dashed (orange long dashed) lines show the pdf's of the longitudinal (transverse)
      components, while the red solid lines is the sum of the two. The distortion due to
      acceptance effects on the right side of the $\cos \theta_{H,\Kstar}$ distribution is
      clearly visible.
      \label{fig:helicity_examples}}
  \end{center}
\end{figure}

The distributions of $\cos \theta_{H, \phi}$ and (in a higher measure) $\cos \theta_{H, \Kstar}$ are
distorted from the ideal theoretical probability density functions (pdf's) by effects related to
the non-uniform acceptance of the detector and other selection criteria. The events with
$\cos \theta_{H, \Kstar} > 0.9$ are particularly affected by these kinds of effects and are
therefore discarded for the final fit. Figure~\ref{fig:helicity_examples} shows the expected
distributions for these variables, for correctly reconstructed signal events for the hypothesis
$f_L = 0.5$.

For each decay mode, we accept at most one signal candidate per event. If an event contains
more than one candidate (which happens very rarely for $B \to \phi K$ and $\sim 10\%$ of the times
for $B \to \phi \Kstar$), we retain the candidate with highest vertex probability for the signal
$B$ candidate. We verify in the simulation that this choice significantly improves the purity
of the sample.

\section{Maximum likelihood fit}
\label{sec:MLfit}

The extraction of the quantities of interest is performed using an unbinned multivariate
maximum likelihood (ML) fit. For the $i^{th}$ input event, the likelihood $\mathcal{L}_i$
is defined as:
\begin{equation}
\mathcal{L}_i = \sum_{j=1}^{m} n_j \mathcal{P}_j(\mbox{\textbf{x}}_i) \; ,
\end{equation}
where $\mathcal{P}_j$ is the probability for the hypothesis (component) $j$ evaluated
for the input variables \textbf{x}$_{i}$, and $n_j$ is the number of events in the whole
sample for the component $j$ ($m$ being the total number of components considered in the fit).
The probability $\mathcal{P}_j$ is the product of the one-dimensional probability
density functions for each of the observables (input variables). One of the main
assumptions of the analysis is that the correlations among these variables
are negligible.

For $N$ input events, the overall likelihood $\mathcal{L}$ is:
\begin{equation}
\mathcal{L} = \frac{e^{-(\sum n_j)}}{N!} \prod_{i=1}^{N} \mathcal{L}_i \; ,
\end{equation}
where the first term takes into account the Poisson fluctuations in the total number of
events.

The input variables entering the ML fit are:
\begin{enumerate}
\item{} $M_{\rm bc}$;
\item{} $\Delta E$;
\item{} $C_{\rm out}^{\prime}$ (the transformed output of the continuum suppression multivariate discriminator $C_{\rm out}$);
\item{} $m(\Kp\Km)$ (invariant mass of the $\phi$ candidate);
\item{} $\cos \theta_{H, \phi}$ (cosine of the helicity angle of the $\phi$ candidate);
\item{} $m(\Kp\pi)$ (invariant mass of the $\Kstar$ candidate);
\item{} $\cos \theta_{H, \Kstar}$ (cosine of the helicity angle of the $\Kstar$ candidate).
\end{enumerate}
The last two variables are relevant only for the $B \to \phi \Kstar$ modes.

The components considered in the fit are:
\begin{itemize}
\item{} \textbf{correctly reconstructed signal events}. For the $B \to \phi \Kstar$ analysis, we
  float separately the yields of the longitudinal ($N_L$) and transverse ($N_T$) components, and compute $f_L$
  taking into account the different reconstruction efficiencies $\varepsilon_L$ and $\varepsilon_T$
  for the longitudinally and transversely polarized events, respectively:
  \begin{equation}
    f_L = \frac{N_L/\varepsilon_L}{N_L/\varepsilon_L + N_T/\varepsilon_T} \, .
  \end{equation}
  The yield parameters are allowed to take negative values (thus the result on $f_L$ may be outside the
  physical $[0, 1]$ range);
\item{} \textbf{self-crossfeed (SXF)}. This component is constituted of signal events in which one
  or more candidate particles originate from the unreconstructed $B$. For the $B \to \phi K$
  analyses, the fraction of self-crossfeed events is negligible, so this component is not considered;
\item{} \textbf{nonresonant}, given by $B \to \Kp \Km K^{(*)}$ events. Early BaBar~\cite{Lees:2012kxa}
  and Belle~\cite{Nakahama:2010nj} analyses have shown that this component accounts for $\mathcal(10\%)$ of
  the events observed in the $\phi$ mass region; this justifies a separate treatment for this
  category of events;
\item{} \textbf{other \BB\ backgrounds};
\item{} \textbf{continuum background}, by far the dominant source of background for this analysis.
\end{itemize}

The continuum background is modeled from the data, excluding the \emph{signal box} defined by
the requirements $M_{\rm bc} > 5.27$ \gevcc\ and $|\Delta E| < 0.1$ \gev. The pdf's of all the other
components are determined from the simulation (\cite{Agostinelli:2002hh}, \cite{Lange:2001uf}).

In the nominal fit, we allow the yields of the signal component(s) and of the continuum background,
to vary, along with the following parameters describing the shape of the continuum background:
the slope of the Argus function~\cite{Albrecht:1990am} that is used to fit the
$M_{\rm bc}$ distribution; the slopes of the (non peaking) $\Delta E$, $m(\Kp\Km)$, and $m(\Kp\pi)$
components; the fractions of the peaking components in the $m(\Kp\Km)$ and $m(\Kp\pi)$ distributions;
and the mean of the core Gaussian component of the $C_{\rm out}^{\prime}$ variable.

The shapes and normalization of the SXF, nonresonant, and other \BB\ background components are
fixed to the expectations from the simulation. The yield of the nonresonant component is fixed to 10\% of the
(total) signal yield; the fraction of the SXF component relative to the correctly reconstructed
signal is kept constant to the predictions of the simulation; and the yield of the other \BB\ backgrounds
is set to the value predicted by the generic Monte Carlo. All these quantities are varied separately by
$\pm 50\%$ for the determination of the systematic uncertainties.

The fitting procedure has been tested extensively using toy Monte Carlo experiments that preserve
the correlations (if any) among the input variables and thus test the main assumption of the fit
model, which assumes all correlations to be negligible. No significant bias has been detected.

The events in the \emph{signal box} have not been analyzed until the fit procedure was clearly defined,
and full confidence was reached from studies on the simulation and data sidebands.

\section{Results}
\label{sec:results}

Tables~\ref{tab:results_VP} and \ref{tab:results_VV} summarize the results of the ML fit applied
to the Summer 2020 Belle II dataset. In all cases, we see a significant signal, whose significance
(taking only into account the statistical uncertainties) ranges from 6.4 to 11.5 standard deviations.
The longitudinal polarization fraction in the $B \to \phi \Kstar$ modes $f_L$ is very consistent
with the expectations.
The branching ratios have been computed using the formula:
\begin{equation}
\mathcal{B}(B \to X) = \frac{N_{sig}}{N(\BB) \times 2 \times \varepsilon \times{\rm ProdBF}} \, ,
\end{equation}
where $N_{sig}$ is the number of fitted signal events, $N(\BB)$ is the number of (charged or neutral)
\BB\ pairs, $\varepsilon$ is the signal reconstruction efficiency, and ProdBF is the product of
the branching fractions of all the intermediate resonances involved. For the $B \to \phi \Kstar$
modes, the branching ratio is the sum of the partial branching ratios for the longitudinal and
transverse components, which have different reconstruction efficiencies.

\begin{table}[htbp]
\begin{center}
  \caption{Summary of the fit results of the $B \to \phi K$ modes.
    The upper part shows the information about the yields (with statistical uncertainty only) of the
    signal (\texttt{nSig}), SXF (\texttt{nSXF}), nonresonant (\texttt{nNR}), and other \BB\ backgrounds
    (\texttt{nBBbar}) components.
    The bottom part displays the reconstruction efficiencies and the measured branching fractions.    
    \label{tab:results_VP}}
  \begin{tabular}{lc@{\hskip 5mm}c}
    \hline\hline
     & $\Bp \to \phi \Kp$ & $\Bz \to \phi \KS$ \\
    \hline
    Events in fit   & 1576 & 695 \\
    \texttt{nSig}   & $55.0 \pm 8.9$ & $15.7 \pm 4.9$ \\
    \texttt{nSXF}   & 0.0 (fixed) & 0.0 (fixed) \\
    \texttt{nNR}    & 5.4 (fixed) & 1.6 (fixed) \\
    \texttt{nBBbar} & 13.0 (fixed) & 3.4 (fixed) \\
    Significance (stat. only) & 11.5$\sigma$ & 6.4$\sigma$ \\
    \hline
    $\varepsilon$ (\%)        & $42.5 \pm 3.0$ & $41.9 \pm 4.8$\\
    $\mathcal{B} (\times 10^{-6})$ & $6.7 \pm 1.1 \pm 0.5$ & $3.0 \pm 0.9 \pm 0.4$ \\
    \hline\hline
  \end{tabular}
\end{center}
\end{table}

\begin{table}[htbp]
\begin{center}
  \caption{Summary of the fit results of the $B \to \phi \Kstar$ modes.
    The upper part shows the information about the yields (with statistical uncertainty only) of the
    longitudinally polarized signal (\texttt{nSigL}), transversely polarized signal (\texttt{nSigT}),
    SXF (\texttt{nSXF}), nonresonant (\texttt{nNR}), and other \BB\ backgrounds (\texttt{nBBbar}) components.
    The bottom part displays the reconstruction efficiencies, the measured branching fractions,
    and longitudinal polarization fractions $f_L$.    
    \label{tab:results_VV}}
  \begin{tabular}{lc@{\hskip 5mm}c}
    \hline\hline
     & $\Bp \to \phi \Kstarp$ & $\Bz \to \phi \Kstarz$ \\
    \hline
    Events in fit    & 2133 & 3055 \\
    \texttt{nSigL}   & $17.6 \pm 5.7$ & $25.0 \pm 7.0$ \\
    \texttt{nSigT}   & $15.2 \pm 5.5$ & $22.7 \pm 7.1$ \\
    \texttt{nSXF}   & 3.7 (fixed) & 4.8 (fixed) \\
    \texttt{nNR}    & 3.3 (fixed) & 4.7 (fixed) \\
    \texttt{nBBbar} & 22.6 (fixed) & 38.2 (fixed) \\
    Significance (stat. only) & $6.4\sigma$ & $9.8\sigma$ \\
    \hline
    $\varepsilon_L$ (\%)        & $31.4 \pm 2.5$ & $32.7 \pm 1.9$\\
    $\varepsilon_T$ (\%)        & $36.8 \pm 2.9$ & $38.6 \pm 2.3$\\
    $\mathcal{B} (\times 10^{-6})$ & $21.7 \pm  4.6 \pm 1.9$ & $11.0 \pm 2.1 \pm 1.1$ \\
    \hline
    $f_L$           & $0.58 \pm 0.23 \pm 0.02$ & $0.57 \pm 0.20 \pm 0.04$ \\
    \hline\hline
  \end{tabular}
\end{center}
\end{table}

In general, the results are in good agreement with the world averages~\cite{Tanabashi:2018oca}.
The observed branching fraction of $B \to \phi \Kstar$ is approximately two standard deviations
higher than the average of the previous results. We checked the stability of the fit by
discarding in turn one of the input variables: in all cases the variations of the signal yield are
less than two events, much smaller than the statistical uncertainty. We also perform a test
in which we remove both helicity angles (so that we lose sensitivity to the polarization),
and again the fitted yield is quite compatible with the nominal fit. We conclude that the fit
is stable, and the higher than expected branching ratio is probably due to a statistical
fluctuation.

\begin{figure}[htbp]
\begin{center}
\includegraphics[width=12cm]{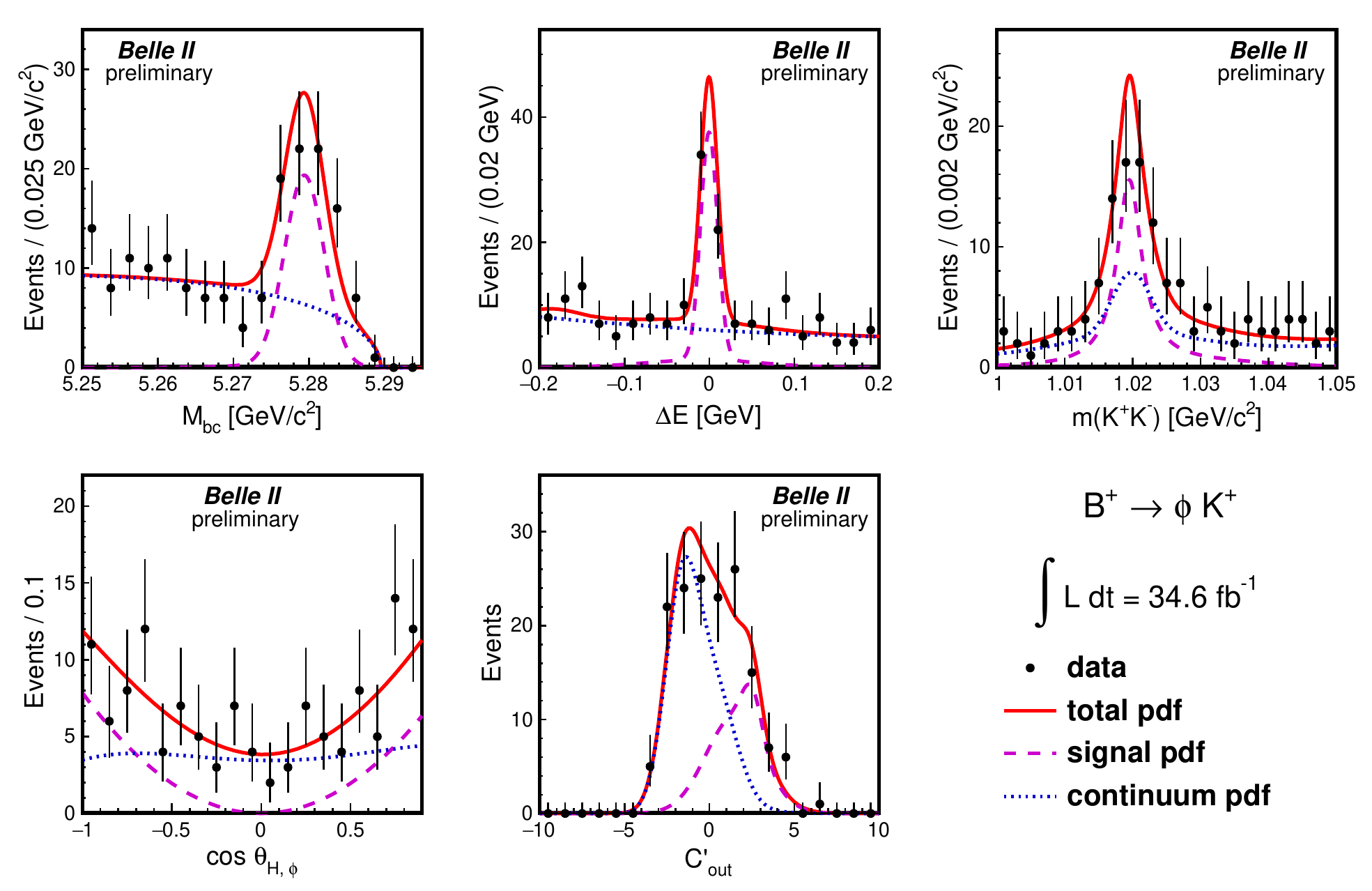}
\caption{Signal enhanced projection plots for the fit variables of
  $\Bp \to \phi \Kp$. The solid red curve represents the total fit function, the
  magenta dashed line shows the signal component, and the blue dotted line
  corresponds to the continuum background.
  \label{fig:PhiKp_ICHEP_projPlots}}
\end{center}
\end{figure}

\begin{figure}[htbp]
\begin{center}
\includegraphics[width=16cm]{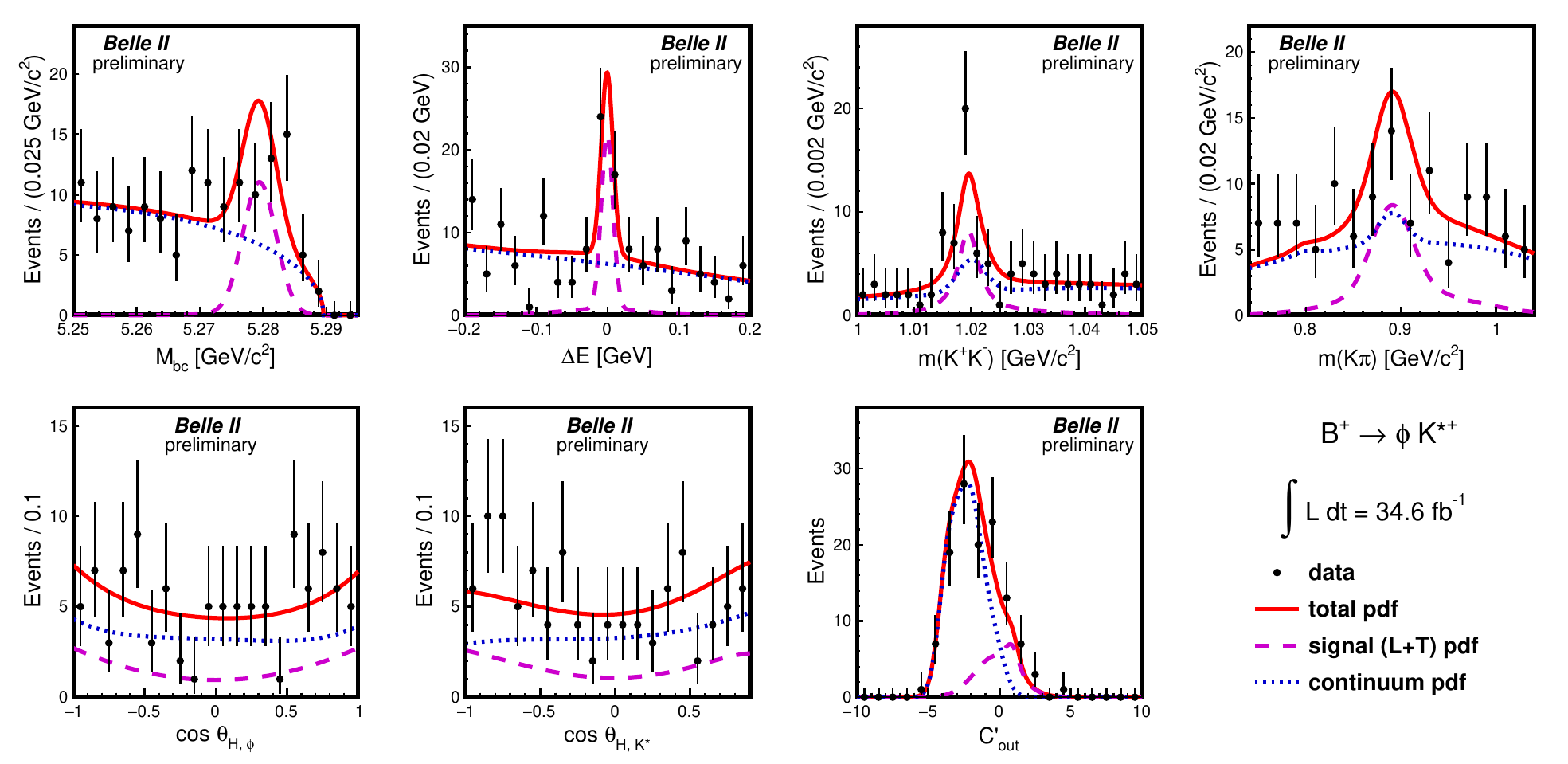}
\caption{Signal enhanced projection plots for the fit variables of
  $\Bp \to \phi \Kstarp$. The solid red curve represents the total fit function, the
  magenta dashed line shows the signal component, and the blue dotted line
  corresponds to the continuum background.
  \label{fig:PhiKstp_ICHEP_projPlots}}
\end{center}
\end{figure}

\begin{figure}[htbp]
\begin{center}
\includegraphics[width=12cm]{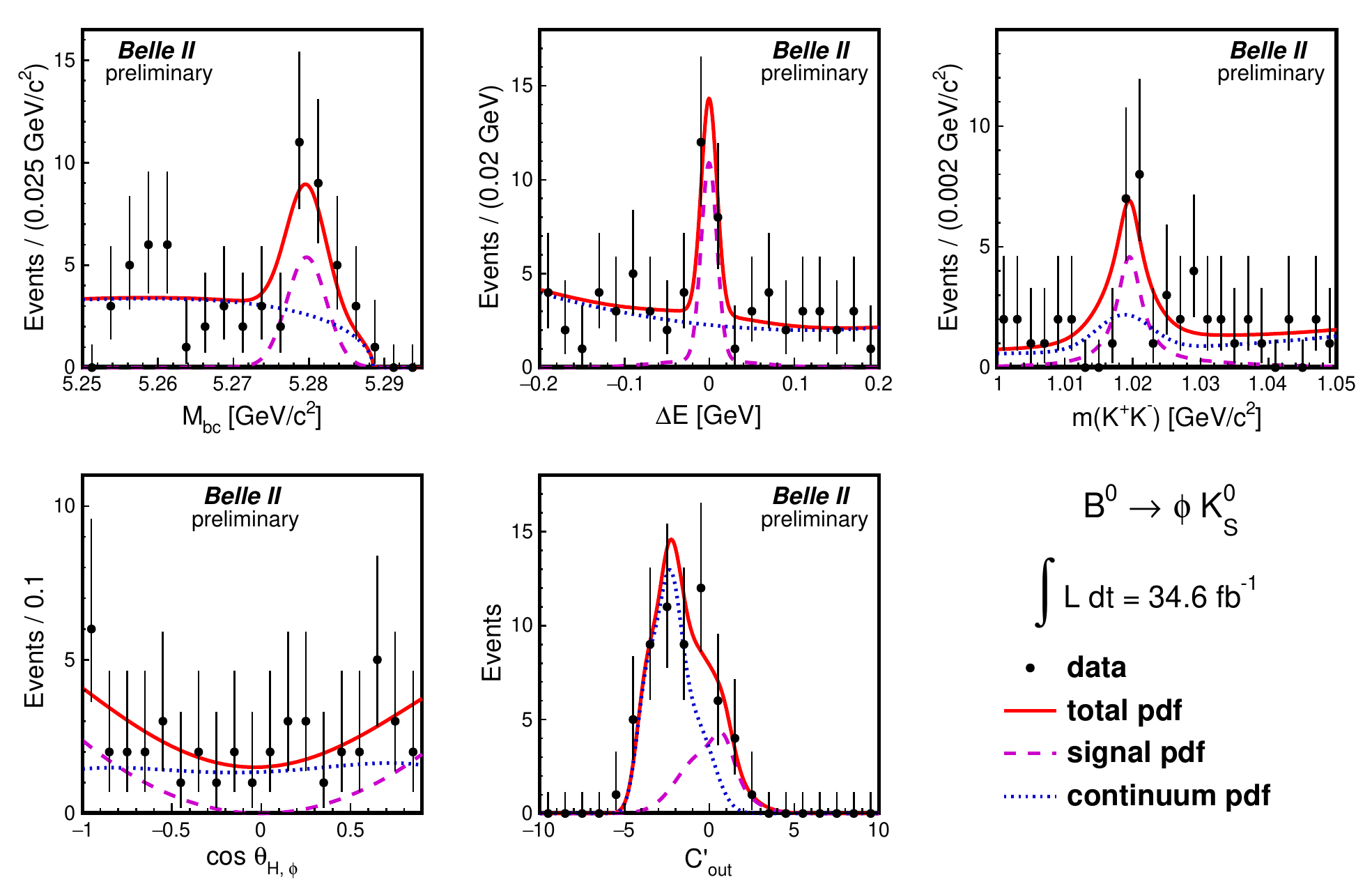}
\caption{Signal enhanced projection plots for the fit variables of
  $\Bz \to \phi \KS$. The solid red curve represents the total fit function, the
  magenta dashed line shows the signal component, and the blue dotted line
  corresponds to the continuum background.
  \label{fig:PhiKs_ICHEP_projPlots}}
\end{center}
\end{figure}

\begin{figure}[htbp]
\begin{center}
\includegraphics[width=16cm]{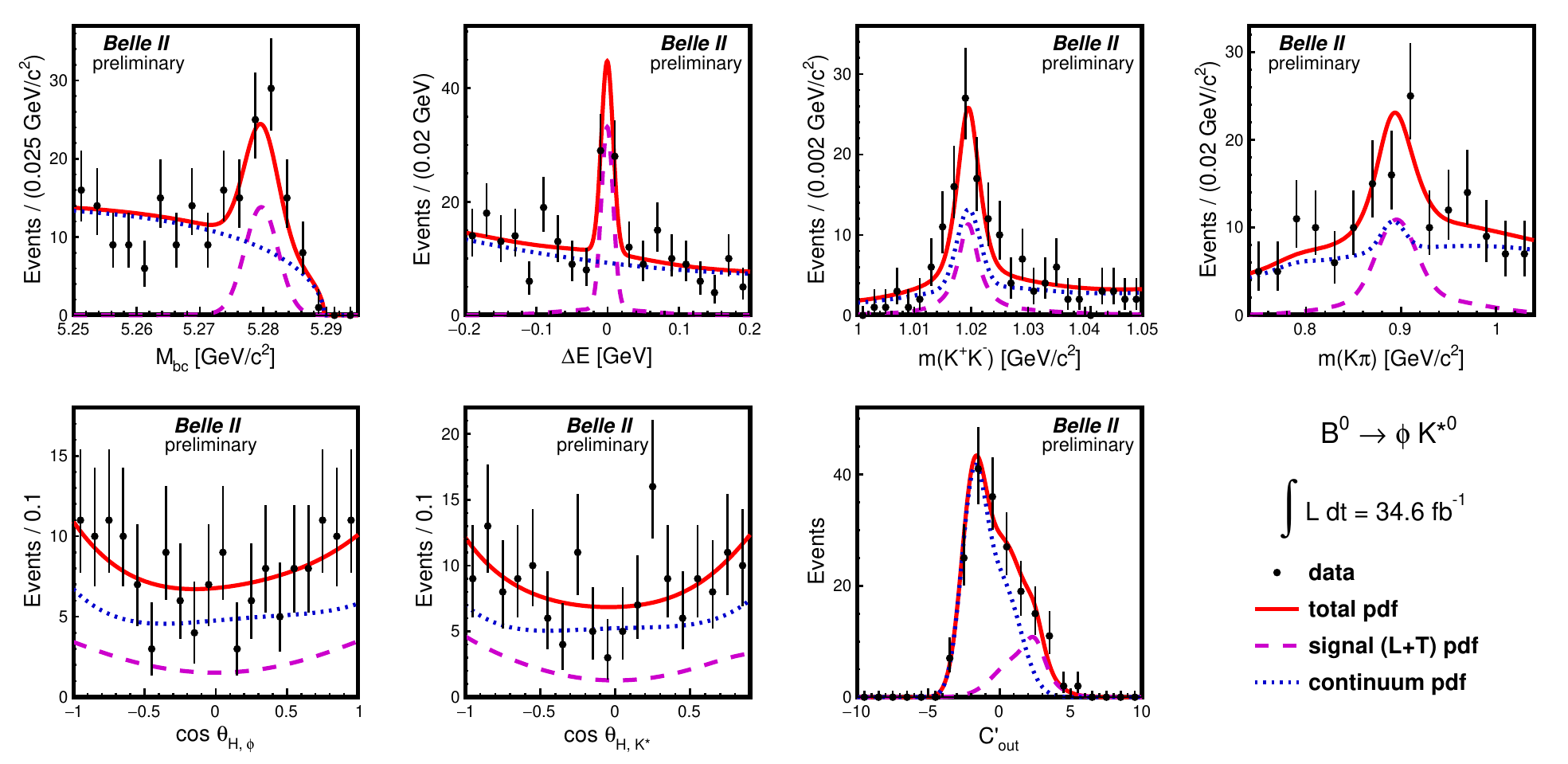}
\caption{Signal enhanced projection plots for the fit variables of
  $\Bz \to \phi \Kstarz$. The solid red curve represents the total fit function, the
  magenta dashed line shows the signal component, and the blue dotted line
  corresponds to the continuum background.
  \label{fig:PhiKst0_ICHEP_projPlots}}
\end{center}
\end{figure}

Figures~\ref{fig:PhiKp_ICHEP_projPlots}--\ref{fig:PhiKst0_ICHEP_projPlots} show the projection
plots of the fit variables utilized for each channel. In order to enhance the signal component,
a cut on the likelihood ratio (for signal over all the hypotheses, with the likelihood being
computed using all the variables with the exception of the one plotted) at 0.5 has been applied.

Finally, we evaluate the compatibility of our results for $f_L$ with the extreme
hypotheses $f_L = 0$ and $f_L = 1$. To do this, we respectively fix to 0 the yield
of the longitudinal and transverse component, and compute $\sqrt{-2 (\log \mathcal{L} - \log \mathcal{L}_0)}$,
where $\mathcal{L}$ is the value of the likelihood computed for the tested hypothesis, and
$\mathcal{L}_0$ is the likelihood of the nominal fit. The lowest significance ($\sim 4.6\sigma$)
is observed for $f_L = 1$ in the $\Bp \to \phi \Kstarp$ channel; in all other cases, the
significance exceeds $5\sigma$.

\section{Systematics}
\label{sec:systematics}

Tables~\ref{tab:syst_summary} and \ref{tab:syst_summary_fL} summarize the systematic uncertainties
affecting the measurements of the branching ratios and of $f_L$, respectively.

\begin{table}[htbp]
\begin{center}
  \caption{Summary of the systematic uncertainties, in per cent, affecting the signal yields.
    The uncertainties are categorized as multiplicative (M) or additive (A).
    \label{tab:syst_summary}}
  \begin{tabular}{lc@{\hskip 5mm}c@{\hskip 5mm}c@{\hskip 5mm}c}
    \hline\hline
    Source & $\Bp \to \phi \Kp$ & $\Bp \to \phi \Kstarp$ & $\Bz \to \phi \KS$ & $\Bz \to \phi \Kstarz$ \\
    \hline
    Tracking efficiency (M)                  & 2.7 & 4.6 & 3.6 & 3.6 \\
    \KS\ reconstruction efficiency (M)       & --  & 6.3 & 10.8 & -- \\
    Kaon ID efficiency (M)                   & 6.4 & 1.1 & 1.0 & 4.7 \\
    Number of \BB\ events (M)                & 2.7 & 2.7 & 2.7 & 2.7 \\
    Modeling of $C_{\rm out^{\prime}}$ (A)   & 1.3 & 1.2 & 1.0 & 5.9 \\
    \BB\ background yield (A)                & 0.3 & 1.2 & 1.4 & 2.3 \\
    Nonresonant yield (A)                    & 3.1 & 1.8 & 4.5 & 3.2 \\
    SXF fraction (A)                         & --  & 0.6 & --  & 1.0 \\
    \hline
    Total multiplicative                     & 7.5 & 8.3 & 11.7 & 6.5 \\
    Total additive                           & 3.4 & 2.5 & 4.8 & 7.1 \\
    \hline
    Total                                    & 8.2 & 8.7 & 12.7 & 9.7 \\
    \hline\hline
  \end{tabular}
\end{center}
\end{table}

\begin{table}[htbp]
\begin{center}
  \caption{Summary of the systematic uncertainties (expressed in absolute values) affecting the
    measurement of $f_L$ in the $B \to \phi \Kstar$ modes.
    \label{tab:syst_summary_fL}}
  \begin{tabular}{lc@{\hskip 5mm}c}
    \hline\hline
    Source & $\Bp \to \phi \Kstarp$ & $\Bz \to \phi \Kstarz$ \\
    \hline
    Acceptance function                   & 0.014 & 0.007 \\
    Modeling of $C_{\rm out^{\prime}}$    & 0.001 & 0.035 \\
    \BB\ background yield                 & 0.002 & 0.009 \\
    Nonresonant yield                     & 0.006 & 0.008 \\
    SXF fraction                          & 0.001 & 0.003 \\
    \hline
    Total                                 & 0.015 & 0.038 \\
    \hline\hline
  \end{tabular}
\end{center}
\end{table}

We consider the following sources of systematics:
\begin{itemize}
\item{} \textbf{tracking efficiency}: we (linearly) add 0.91\% for each charged track
  in the signal final state;
\item{} \textbf{\KS\ reconstruction efficiency}: we use a data control sample, and we
  observe that the \KS\ reconstruction efficiency decreases (compared to the simulation)
  linearly with the flight length. We apply an uncertainty of 1\% for each cm of average flight
  length of the \KS\ candidate;
\item{} \textbf{charged kaon identification}: we take the difference between the reconstruction
  efficiency for signal candidates measured using only Monte Carlo information and the efficiency
  obtained by shifting the kaon identification efficiency to what is measured on a data sample
  of $\Dstarp \to \pip D^0 (\to \Km \pip)$. This uncertainty is larger for the $\phi \Kp$ and
  $\phi \Kstarz$ mode, as the charged kaon typically has a much higher momentum than the kaons
  produced by the $\phi$ decay, and the agreement between data and simulation is currently much
  better at lower momenta;
\item{} \textbf{number of \BB\ events}: we assign a 2.7\% systematic error, which includes the
  uncertainty on cross-section, integrated luminosity, and potential shifts from the peak
  CM energy during the run periods;
\item{} \textbf{modeling of the $C_{\rm out^{\prime}}$ variable}: we take the difference in the
  results obtained when the shape of the $C_{\rm out^{\prime}}$ is determined from the data sidebands
  (nominal fit) and when the shape is modeled from the simulation;
\item{} \textbf{yields of SXF, nonresonant, and \BB\ background components}: we individually
  vary by $\pm 50\%$ the yield of each component, and take the difference of the results
  (with respect to the nominal fit) as systematic uncertainty;
\item{} \textbf{acceptance function for the helicity angles} (relevant only for the measurement
  of $f_L$). The pdf's of $\cos \theta_{H, \phi}$ and $\cos \theta_{H, \Kstar}$ are the products
  of a theoretical pdf's and an acceptance function. We evaluate the systematic uncertainty
  by taking the difference between the nominal fit results and the cases in which the deviations
  from unity of the acceptance function are doubled or removed (uniform acceptance).
\end{itemize}

\section{Conclusions}
\label{sec:conclusions}

In conclusion, we have observed all four $B \to \phi K^{(*)}$ channels in the
Summer 2020 Belle II dataset of 34.6\invfb, with branching ratios that are in good or fair (for the
$\Bp \to \phi \Kstarp$ case) agreement with the world averages \cite{Tanabashi:2018oca}.
The measurement of the longitudinal polarization fraction $f_L$ is in excellent
agreement with our expectations.

The results of this analysis are summarized in Table~\ref{tab:conclusions}. We also
compute the isospin ratios
\begin{equation}
I_{\phi K^{(*)}} = \frac{\mathcal{B}(\Bp \to \phi K^{(*)+})}{\mathcal{B}(\Bz \to \phi K^{(*)0})} \, ,
\end{equation}
which are interesting observables for detecting signs of physics beyond the standard model
(e.g. in~\cite{Feldmann:2008fb}) and that we measure to be in reasonably good agreement with unity.

\begin{table}[htbp]
\begin{center}
  \caption{Summary of the results obtained in this analysis.
    \label{tab:conclusions}}
  \begin{tabular}{lc@{\hskip 5mm}c}
    \hline\hline
     & This analysis & World Average~\cite{Tanabashi:2018oca} \\
    $\mathcal{B} (\times 10^{-6})$ \\
    \hline
    $\phi \Kp$     & $6.7 \pm 1.1 \pm 0.5$  & $8.8 \pm 0.7$ \\
    $\phi \Kz$     & $5.9 \pm 1.8 \pm 0.7$  & $7.3 \pm 0.7$ \\
    $I_{\phi K}$     & $1.1 \pm 0.4 \pm 0.2 $ & $1.21 \pm 0.15$ \\
    \hline
    $\phi \Kstarp$ & $21.7 \pm 4.6 \pm 1.9$ & $10.0 \pm 2.0$ \\
    $\phi \Kstarz$ & $11.0 \pm 2.1 \pm 1.1$ & $10.0 \pm 0.5$ \\
    $I_{\phi K^{*}}$  & $2.0 \pm 0.6 \pm 0.3$ & $1.00 \pm 0.21$ \\
    \hline\hline
    $f_L$ \\
    \hline
    $\phi \Kstarp$ & $0.58 \pm 0.23 \pm 0.02$ & $0.50 \pm 0.05$ \\
    $\phi \Kstarz$ & $0.57 \pm 0.20 \pm 0.04$ & $0.497 \pm 0.017$ \\
    \hline\hline
  \end{tabular}
\end{center}
\end{table}

\section*{Acknowledgements}
We thank the SuperKEKB group for the excellent operation of the
accelerator; the KEK cryogenics group for the efficient
operation of the solenoid; and the KEK computer group for
on-site computing support.
This work was supported by the following funding sources:
Science Committee of the Republic of Armenia Grant No. 18T-1C180;
Australian Research Council and research grant Nos.
DP180102629, 
DP170102389, 
DP170102204, 
DP150103061, 
FT130100303, 
and
FT130100018; 
Austrian Federal Ministry of Education, Science and Research, and
Austrian Science Fund No. P 31361-N36; 
Natural Sciences and Engineering Research Council of Canada, Compute Canada and CANARIE;
Chinese Academy of Sciences and research grant No. QYZDJ-SSW-SLH011,
National Natural Science Foundation of China and research grant Nos.
11521505,
11575017,
11675166,
11761141009,
11705209,
and
11975076,
LiaoNing Revitalization Talents Program under contract No. XLYC1807135,
Shanghai Municipal Science and Technology Committee under contract No. 19ZR1403000,
Shanghai Pujiang Program under Grant No. 18PJ1401000,
and the CAS Center for Excellence in Particle Physics (CCEPP);
the Ministry of Education, Youth and Sports of the Czech Republic under Contract No.~LTT17020 and 
Charles University grants SVV 260448 and GAUK 404316;
European Research Council, 7th Framework PIEF-GA-2013-622527, 
Horizon 2020 Marie Sklodowska-Curie grant agreement No. 700525 `NIOBE,' 
and
Horizon 2020 Marie Sklodowska-Curie RISE project JENNIFER2 grant agreement No. 822070 (European grants);
L'Institut National de Physique Nucl\'{e}aire et de Physique des Particules (IN2P3) du CNRS (France);
BMBF, DFG, HGF, MPG, AvH Foundation, and Deutsche Forschungsgemeinschaft (DFG) under Germany's Excellence Strategy -- EXC2121 ``Quantum Universe''' -- 390833306 (Germany);
Department of Atomic Energy and Department of Science and Technology (India);
Israel Science Foundation grant No. 2476/17
and
United States-Israel Binational Science Foundation grant No. 2016113;
Istituto Nazionale di Fisica Nucleare and the research grants BELLE2;
Japan Society for the Promotion of Science,  Grant-in-Aid for Scientific Research grant Nos.
16H03968, 
16H03993, 
16H06492,
16K05323, 
17H01133, 
17H05405, 
18K03621, 
18H03710, 
18H05226,
19H00682, 
26220706,
and
26400255,
the National Institute of Informatics, and Science Information NETwork 5 (SINET5), 
and
the Ministry of Education, Culture, Sports, Science, and Technology (MEXT) of Japan;  
National Research Foundation (NRF) of Korea Grant Nos.
2016R1\-D1A1B\-01010135,
2016R1\-D1A1B\-02012900,
2018R1\-A2B\-3003643,
2018R1\-A6A1A\-06024970,
2018R1\-D1A1B\-07047294,
2019K1\-A3A7A\-09033840,
and
2019R1\-I1A3A\-01058933,
Radiation Science Research Institute,
Foreign Large-size Research Facility Application Supporting project,
the Global Science Experimental Data Hub Center of the Korea Institute of Science and Technology Information
and
KREONET/GLORIAD;
Universiti Malaya RU grant, Akademi Sains Malaysia and Ministry of Education Malaysia;
Frontiers of Science Program contracts
FOINS-296,
CB-221329,
CB-236394,
CB-254409,
and
CB-180023, and SEP-CINVESTAV research grant 237 (Mexico);
the Polish Ministry of Science and Higher Education and the National Science Center;
the Ministry of Science and Higher Education of the Russian Federation,
Agreement 14.W03.31.0026;
University of Tabuk research grants
S-1440-0321, S-0256-1438, and S-0280-1439 (Saudi Arabia);
Slovenian Research Agency and research grant Nos.
J1-9124
and
P1-0135; 
Agencia Estatal de Investigacion, Spain grant Nos.
FPA2014-55613-P
and
FPA2017-84445-P,
and
CIDEGENT/2018/020 of Generalitat Valenciana;
Ministry of Science and Technology and research grant Nos.
MOST106-2112-M-002-005-MY3
and
MOST107-2119-M-002-035-MY3, 
and the Ministry of Education (Taiwan);
Thailand Center of Excellence in Physics;
TUBITAK ULAKBIM (Turkey);
Ministry of Education and Science of Ukraine;
the US National Science Foundation and research grant Nos.
PHY-1807007 
and
PHY-1913789, 
and the US Department of Energy and research grant Nos.
DE-AC06-76RLO1830, 
DE-SC0007983, 
DE-SC0009824, 
DE-SC0009973, 
DE-SC0010073, 
DE-SC0010118, 
DE-SC0010504, 
DE-SC0011784, 
DE-SC0012704; 
and
the National Foundation for Science and Technology Development (NAFOSTED) 
of Vietnam under contract No 103.99-2018.45.

\bibliography{references}

\end{document}